\documentclass[aps,pra,twocolumn,superscriptaddress,showpacs,showkeys,amsmath,amssymb]{revtex4}

\usepackage{amsfonts}
\usepackage{amssymb,amsmath}
\usepackage{mathrsfs}
\usepackage{latexsym}
\usepackage{amsmath}
\usepackage[cp1251]{inputenc}
\usepackage{graphicx}
\usepackage{dcolumn}
\usepackage{bm}
\usepackage{color}

\RequirePackage{ifthen}
\RequirePackage[pdfstartview=FitH]{hyperref}
\begin{document}

    \title{Impurity in a three-dimensional unitary Bose gas}
    \author{O.~Hryhorchak}
    \affiliation{Department for Theoretical Physics, Ivan Franko National University of Lviv, 12 Drahomanov Str., Lviv, Ukraine}
    \author{G.~Panochko}
    \affiliation{Department of Optoelectronics and Information Technologies, Ivan Franko National University of Lviv, 107 Tarnavskyj Str., Lviv, Ukraine}
    \author{V.~Pastukhov\footnote{e-mail: volodyapastukhov@gmail.com}}
    \affiliation{Department for Theoretical Physics, Ivan Franko National University of Lviv, 12 Drahomanov Str., Lviv, Ukraine}

    \date{\today}

    \pacs{67.85.-d}

    \keywords{Bose polaron, unitary Bose gas, local density approximation, non-linear Schr\"odinger equation}
    \begin{abstract}
    	By using simple and efficient method we discuss properties of a single impurity immersed in three-dimensional Bose gas with the interaction between particles tuned to unitary limit. Particularly, adopting the mean-field-like approximation we present the first estimations for the low-momentum parameters of the impurity spectrum, namely, the binding energy, the effective mass and the quasiparticle residue both for repulsive and attractive Bose polarons in the unitary gas. 
   	\end{abstract}

    \maketitle

\section{Introduction}
\label{sec1}
The creation of the Bose-Einstein condensed state in ultra-dilute alkalies \cite{Anderson,Davis} 25 years ago opened up new horizons for studying collective phenomena in these essentially quantum systems. The next step in the experimental progress was achieved by the possibility of controlling the small amount of particles, which provided the observation of Efimov's states \cite{Kraemer} and more recently the so-called Fermi \cite{Schirotzek} and Bose polarons \cite{Jorgensen,Hu,Camargo,Yan}, i.e., a single impurity atoms immersed in majority of host fermions and bosons, respectively. Such a success of experimental techniques provided the verification of numerous theoretical predictions of the Bose polaron properties in three dimension, based on variational \cite{Novikov_10,Li_14}, renormalization-group \cite{Grusdt_15,Grusdt_16}, field-theoretical \cite{Novikov_09,Rath_13} and Monte Carlo (MC) \cite{Vlietinck_15,Pena_Ardila_15,Pena_Ardila_16} methods which allow the description of the quantum effects for weak \cite{Grusdt_Demler_15,Christensen_15,Panochko_17} and strong \cite{Pena_Ardila_19} boson-impurity interactions and were applicable at low temperatures and in the vicinity of the Bose-system superfluid transition point \cite{Boudjemaa,Levinsen_17,Guenther,Bosepolaron_D,Field}.

Along with such a wide range of works where polarons are investigated in diluted Bose gases, the properties of impurities in the strongly-interacting systems are well-studied only in a context of liquid $^4$He \cite{Boronat,Galli,Rossi_2004,Lemeshko,Boninsegni,Panochko_18}. But properties of superfluid helium itself and impurity atoms immersed in it are strongly dependent on the details of the two-body $^4$He-$^4$He potential. There is, however, another experimentally realizable \cite{Rem_et_al,Makotyn_et_al,Fletcher_et_al} gas of bosons with short lifetime and strong point-like inter-particle interactions, the so-called unitary Bose gas. The two-body $s$-wave scattering length formally diverges in this system, and thermodynamics is universal \cite{Ho} in a sense that is fully controlled by single dimensionless parameter, namely, temperature in units of bosonic analog of the Fermi energy. To our knowledge, behavior of exterior particles immersed in unitary Bose gas was not yet explored. The objective of this article is, therefore, partially to fill this gap by utilizing the semi-phenomenological approach of Ref.~\cite{Hryhorchak_et_al} to the problem of impurity in a three-dimensional unitary Bose gas at absolute zero. Such a mean-field-like (MF) treatments \cite{Gross_62,Astrakharchik_04}, regardless of their simplicity, were recently demonstrated \cite{Volosniev_17,Pastukhov_3BIBP,Panochko_19,Smith} to be quite efficient for the description, in reasonable agreement with the MC simulations \cite{Parisi_17,Grusdt_17} of Bose polarons in one dimension. In contrast to one-dimensional systems, where the effects of quantum fluctuations are strongly exhausted \cite{Jager} a behavior of impurities in higher dimensions is known to be much influenced by the few-body physics. Of course, the MF ansatz cannot describe a formation of the boson-impurity dimers, trimers etc. \cite{Yoshida}, but it gives reliable results for the Bose-polaronic state at any strength of the boson-impurity couplings.

\setcounter{equation}{0}

\section{Model and method}
The considered model consists of a single impurity atom immersed in a bosonic medium at absolute zero. The Bose subsystem that contains $N$ spinless particles is described by the semi-phenomenological model with the Hamiltonian that goes back to the seminal work of Landau \cite{Landau}
\begin{eqnarray}\label{H_B}
H_B=\int d{\bf r}\left\{-\frac{\hbar^2}{2m}\psi^+({\bf r})\nabla^2\psi({\bf r})+\mathcal{E}[n({\bf r})]\right\},
\end{eqnarray}
where the integration is carried out over large volume $V$ with periodic boundary conditions, $m$ is the mass of bosons and $\mathcal{E}[n]$ stands for the normal-ordered energy density of the uniform system at rest. The field operators $\psi^+({\bf r})$, $\psi({\bf r})$ obey standard bosonic commutation relations and $n({\bf r})=\psi^+({\bf r})\psi({\bf r})$ denotes the operator of a local bosonic density. The part of total Hamiltonian which is fully microscopical and refers to the impurity, reads
\begin{eqnarray}\label{H_I}
H_I=-\frac{\hbar^2}{2m_I}\frac{\partial^2}{\partial {\bf r}_I^2}+\int d{\bf r}\,\Phi(|{\bf r}_I-{\bf r}|)n({\bf r}),
\end{eqnarray}
where ${\bf r}_I$ and $m_I$ denote a position and a mass of an impurity, respectively. The second term in (\ref{H_I}) is the energy of impurity-Bose-system interaction with the two-body potential $\Phi(r)$ that will be specified below. Our choice of bosonic Hamiltonian is a very flexible one, because with such a form it exactly reproduces bosons with the point-like interaction and allows the phenomenological generalizations, in the spirit of local density approximation, by varying the energy density $\mathcal{E}[n]$.

The whole zero-temperature analysis below is based on the variation principle for total Hamiltonian $H_B+H_I$. As a trial wave function we choose the MF ansatz (here $|\textrm{vac}\rangle $ is the normalized Fock vacuum state)
\begin{eqnarray}\label{Psi}
|\Psi\rangle_N=\frac{\left(b^+_0\right)^N}{\sqrt{N!}}|\textrm{vac}\rangle,
\end{eqnarray}
with the bosonic creation operator defined as follows
\begin{eqnarray}\label{b_0+}
b^+_0=\frac{1}{\sqrt{N}}\int d{\bf r}\,\phi({\bf r}_I-{\bf r})\psi^+({\bf r}),
\end{eqnarray}
where unknown function $\phi({\bf r})$ is the subject for minimization of the energy functional
\begin{eqnarray}\label{E_funct}
E=\frac{1}{V}\int d{\bf r}_I\,{_N}\langle\Psi| H_B+H_I |\Psi\rangle_N,
\end{eqnarray}
with additional constrain $\int d{\bf r}|\phi({\bf r})|^2=N$, which accounts for the correct normalization (we chose the one, where ${_N}\langle\Psi|n({\bf r})|\Psi\rangle_N=|\phi({\bf r}_I-{\bf r})|^2$ is the local density of bosons). It is readily seen that our trial wave function, unlike previous studies \cite{Cucchietti_06,Kalas_06,Sacha_06,Bruderer_08,Roberts_09,Blinova_13}, preserves the continuous translation symmetry of the system. It means, that this ansatz is capable to capture polaronic behavior and cannot provide the description of states with formation of the boson-impurity molecules. It is well-known that the MF  approximation typically misses important aspects of the few-body physics, but can describe the self-localization phenomenon. The latter can be traced back by the enormous increase of the impurity effective mass, but has not been confirmed by the MC simulations \cite{Pena_Ardila_15} of Bose polarons. One more important thing about the ansatz is that the average total momentum of the system provided by the wave function (\ref{Psi}) is equal to zero identically. But slight modification of $|\Psi\rangle_N$, namely
\begin{eqnarray}\label{Psi_p}
|\Psi_{\bf p}\rangle_N=e^{i{\bf p}{\bf r}_I/\hbar}\frac{\left(\tilde{b}^+_0\right)^N}{\sqrt{N!}}|\textrm{vac}\rangle,
\end{eqnarray}
with $\tilde{b}^+_0$ being almost equal to $b^+_0$ except for the replacement $\phi({\bf r}_I-{\bf r})\to \phi_{\bf p}({\bf r}_I-{\bf r})$, can describe the system moving as a whole with finite momentum ${\bf p}$. The expectation value of the impurity momentum then reads
\begin{eqnarray}\label{p_imp}
{\bf p}_{I}={\bf p}-i\hbar\int d{\bf r}\,\phi^*_{\bf p}({\bf r})\nabla\phi_{\bf p}({\bf r}).
\end{eqnarray}
The minimization of the total energy in state $|\Psi_{\bf p}\rangle_N$ should be therefore carried out with the additional constrain on the impurity momentum (\ref{p_imp}). Particularly, fixing the non-zero (\ref{p_imp}) one can describe the impurity moving with finite momentum.
The computation of functional (\ref{E_funct}) is rather trivial if one notices that $\psi({\bf r})|\Psi_{\bf p}\rangle_N=\phi_{\bf p}({\bf r}_I-{\bf r})|\Psi_{\bf p}\rangle_{N-1}$. Then an averaging (in $N\gg 1$ limit) of $H_B$ and the second term in $H_I$ reduces to the replacement of field operators $\psi^+({\bf r})$ and $\psi({\bf r})$ by $\phi^*_{\bf p}({\bf r}_I-{\bf r})$ and $\phi_{\bf p}({\bf r}_I-{\bf r})$, respectively. The appropriate calculations for the impurity kinetic energy operator yield 
\begin{eqnarray}\label{E_I}
\frac{{\bf p}^2_{I}}{2m_I}
+\frac{\hbar^2}{2m_I}\int d{\bf r}\,|\nabla\phi_{\bf p}({\bf r})|^2.
\end{eqnarray}
Combining everything together, we have
\begin{eqnarray}\label{E_explicit}
	E=\frac{{\bf p}^2_{I}}{2m_I}+\int d{\bf r}\left\{\frac{\hbar^2}{2m_r}\left|\nabla \phi_{\bf p}({\bf r})\right|^2\right.\nonumber\\
	\left.+\Phi(r)|\phi_{\bf p}({\bf r})|^2+\mathcal{E}\left[|\phi_{\bf p}({\bf r})|^2\right]\right\},
\end{eqnarray}
where $m_r=m_Im/(m_I+m)$ is the boson-impurity reduced mass. Minimization of the above functional with constrain on ${\bf p}_I$ leads to the complicated non-linear Schr\"odinger equation \cite{Hryhorchak_et_al}. At zero impurity momentum, however, this equation substantially simplifies
\begin{align}\label{phi}
	-\frac{\hbar^2\nabla^2}{2m_r}\phi({\bf r})
	+\Phi(r)\phi({\bf r})+\mathcal{E}'[|\phi({\bf r})|^2]\phi({\bf r})=\mu\phi({\bf r}),
\end{align}
where prime near $\mathcal{E}[n]$ denotes the derivative with respect to density $n$, the Lagrange multiplier $\mu$ accounts for the proper normalization of $\phi({\bf r})$ which, in turn, can be chosen as the real-valued function. The later constrain automatically provides the vanishing value of the average impurity momentum. With the solution of Eq.~(\ref{phi}) in hands we can calculate the total energy $E$ of the system `impurity+bosons'. Then, by subtracting the energy of `pure' bosons $V\mathcal{E}[n]$ (here $n=N/V$ is the uniform density of Bose particles without an impurity) we obtain the Bose polaron binding energy $\varepsilon_I$.

Although, the function that extremizes functional (\ref{E_explicit}) is very complicated, the general structure of spectrum of the slowly moving polaron is known
\begin{eqnarray}\label{varepsilon_p}
	\varepsilon_I(p)=\varepsilon_I+\frac{p^2}{2m^*_I}+\mathcal{O}(p^4),
\end{eqnarray}
where the effective mass $m^*_I$ is introduced. It can be shown from general arguments that $m^*_I$ is always larger than bare mass of the impurity atom both for repulsive and attractive boson-impurity potentials. Being the first correction to the Bose polaron spectrum at small ${\bf p}$, the effective mass is fully determined \cite{Hryhorchak_et_al}, in thermodynamic limit, by the density profile of bosons in presence of \textit{motionless} impurity
\begin{eqnarray}\label{m*_I}
\frac{m^*_I}{m_I}=1+\frac{m_r}{m_I}\int d{\bf r}\frac{\left[|\phi({\bf r})|^2-|\phi_{\infty}|^2\right]^2}{2|\phi_{\infty}|^2+|\phi({\bf r})|^2},
\end{eqnarray}
where $\phi_{\infty}=\phi({\bf r}\to \infty)$. Therefore we only have to obtain function $\phi({\bf r})$ for motionless impurity in order to calculate two parameters in the r.h.s. of Eq.~(\ref{varepsilon_p}) of the low-momentum Bose polaron spectrum.
  
For completeness, we give the general formula \cite{Hryhorchak_et_al} of the MF quasiparticle residue
\begin{eqnarray}\label{Z}
Z=\exp \left\{-\int d{\bf r}\left(|\phi({\bf r})|-|\phi_{\infty}|\right)^2\right\},
\end{eqnarray}
which is determined by the modulus squared of the wave-functions overlap with and without impurity immersed. Note that Eq.~(\ref{Z}) is applicable only in a case of motionless ${\bf p}=0$ impurity.

\section{Results}
Before proceeding with the discussion of our main results concerning a behavior of the impurity in unitary Bose gas, it is instructive to test this approach on the Bose polaron immersed in a dilute gas. The case of repulsive $\Phi(r)$ has been already reported \cite{Hryhorchak_et_al}, therefore, here we mainly focus on the attractive branch.

\subsection{Attractive Bose polaron in dilute gas}
In the extremely dilute limit the energy density of bosons is well-described by the MF expression $\mathcal{E}[n]=\frac{2\pi\hbar^2a}{m}n^2$, where $a$ is the two-boson $s$-wave scattering length. In order to model an attractive boson-impurity interaction we have chosen the spherical potential well
\begin{eqnarray}\label{Phi}
\Phi(r)=\left\{\begin{array}{c}
-\Phi_0, \ \ r\le R\\
0, \ \  r>R
\end{array}\right.,
\end{eqnarray} 
with finite range $R$. The depth of potential $\Phi_0$ can be easily related \cite{Landau_QM} through the transcendental equation
\begin{eqnarray}\label{a_I}
a_I=R\left\{1-\frac{\tan(k_0R)}{k_0R}\right\}, \ \ k_0=\frac{\sqrt{2m_r\Phi_0}}{\hbar},
\end{eqnarray} 
to the boson-impurity $s$-wave scattering length $a_I$ and range $R$. The numerical solution of Eq.~(\ref{phi}) (with $\Phi(r)$ and $\mathcal{E}[n]$ inserted) was performed for dimensionless boson-boson interaction $na^3=10^{-5}$, equal masses of particles $m_I=m$, negative values of $a_I$ and at few ranges of the potential well $R=a$, $R=a/5$, $R=a/10$. For comparison with MC simulations \cite{Pena_Ardila_15}, one should follow value $R=a/5$. In fact, equilibrium properties of the impurity is almost unaffected by a precise value of this parameter and results for a smaller $R$s only slightly differ but require much more computer time. Note also that choice of $\Phi(r)$ allows the unitary limit, where formally $a_I\to -\infty$. 
\begin{figure}[h!]
	\includegraphics[width=0.45\textwidth,clip,angle=-0]{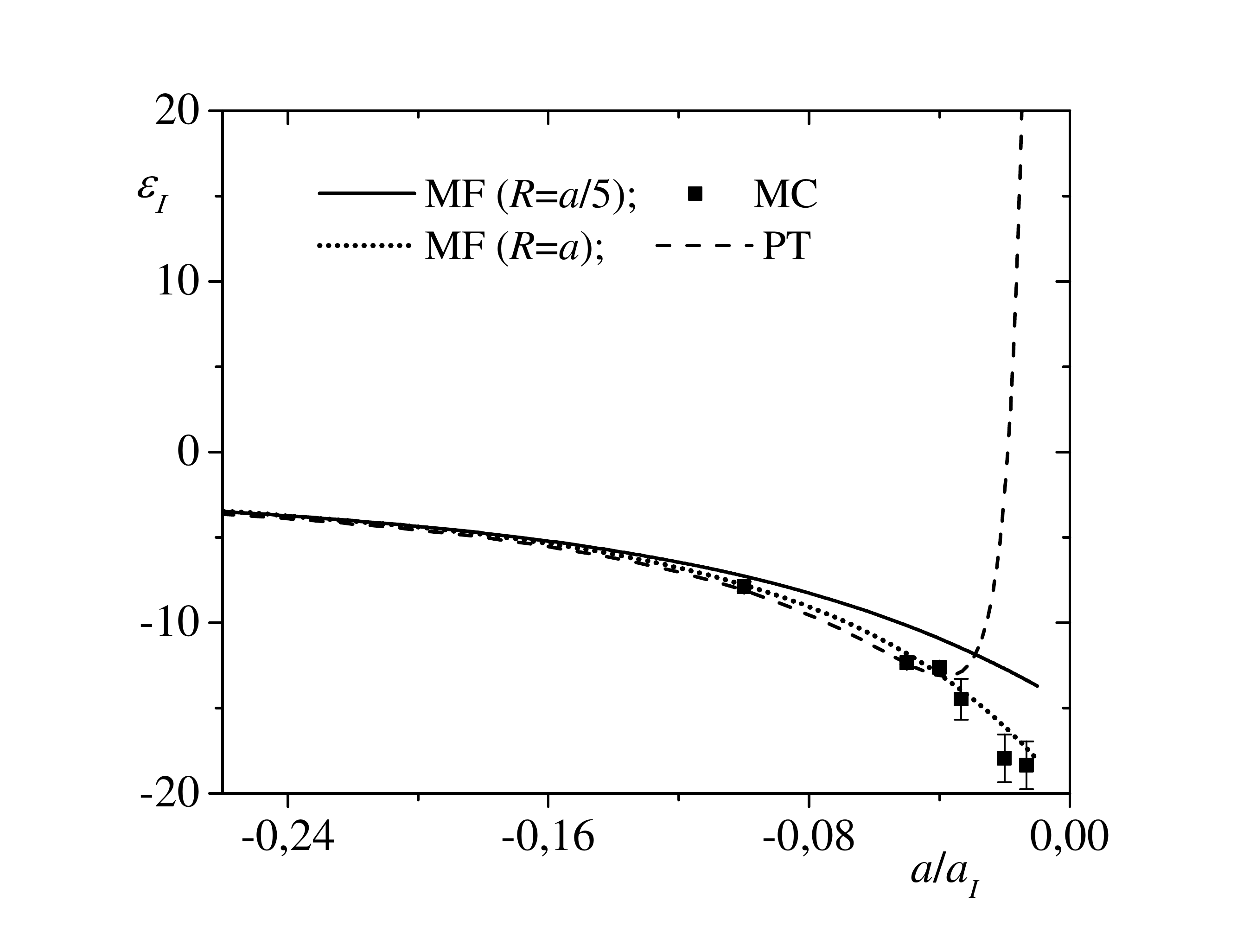}
	\caption{Impurity binding energy (in units of the MF chemical potential $\mu=\frac{4\pi\hbar^2a}{m}n$ of bosons) in dilute Bose gas, $na^3=10^{-5}$. Solid and dotted lines are our MF result at $R<a/5$ and $R<a$, respectively. Result at $R<a/10$ is almost indistinguishable from the one at $R<a/5$, therefore, not plotted. Dashed line displays the perturbative calculations up to the second order \cite{Grusdt_Demler_15,Christensen_15,Panochko_17}. Symbols denote results of the MC simulations \cite{Pena_Ardila_15}.}
	\label{energy_Mf}
\end{figure}
\begin{figure}[h!]
	\includegraphics[width=0.45\textwidth,clip,angle=-0]{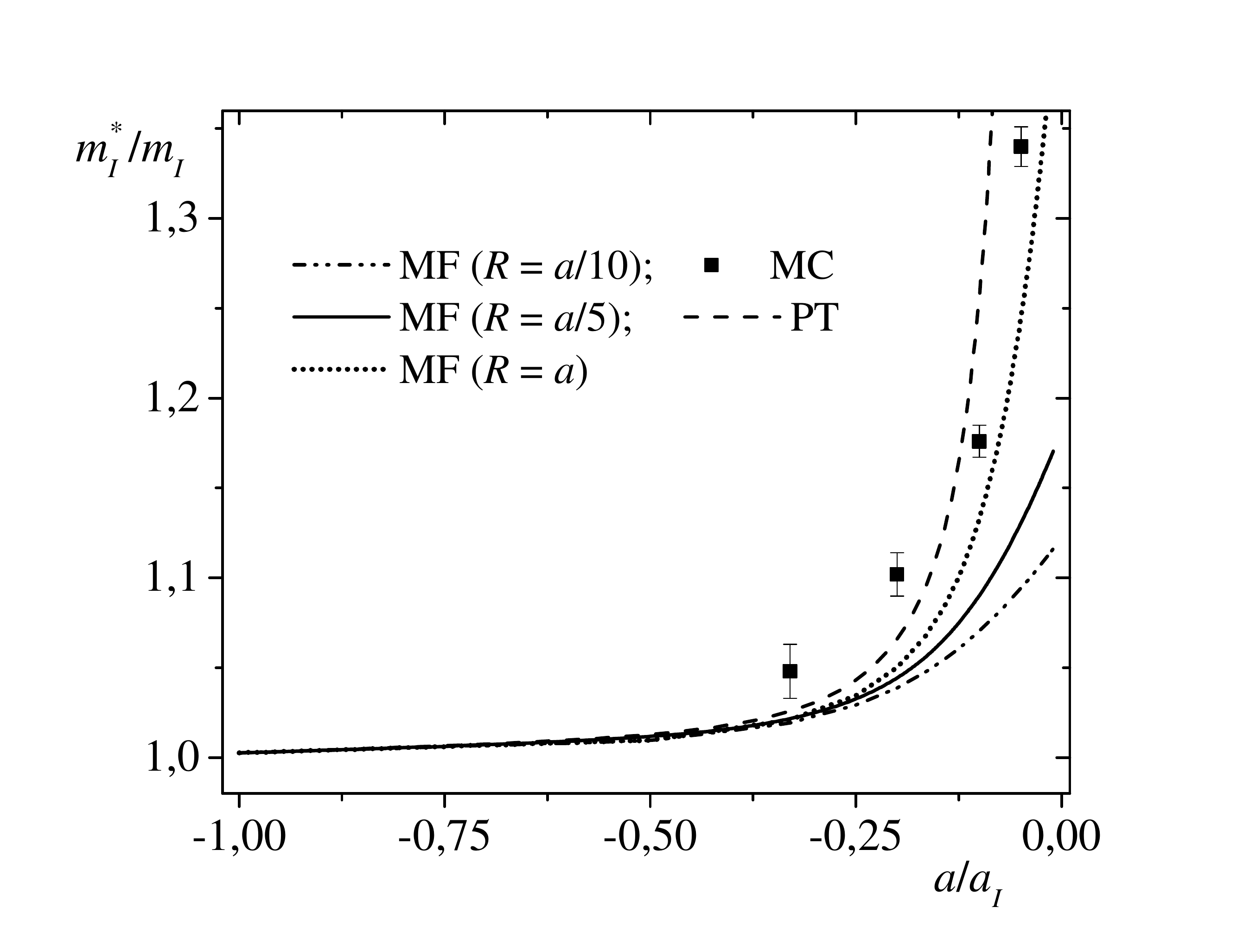}
	\includegraphics[width=0.45\textwidth,clip,angle=-0]{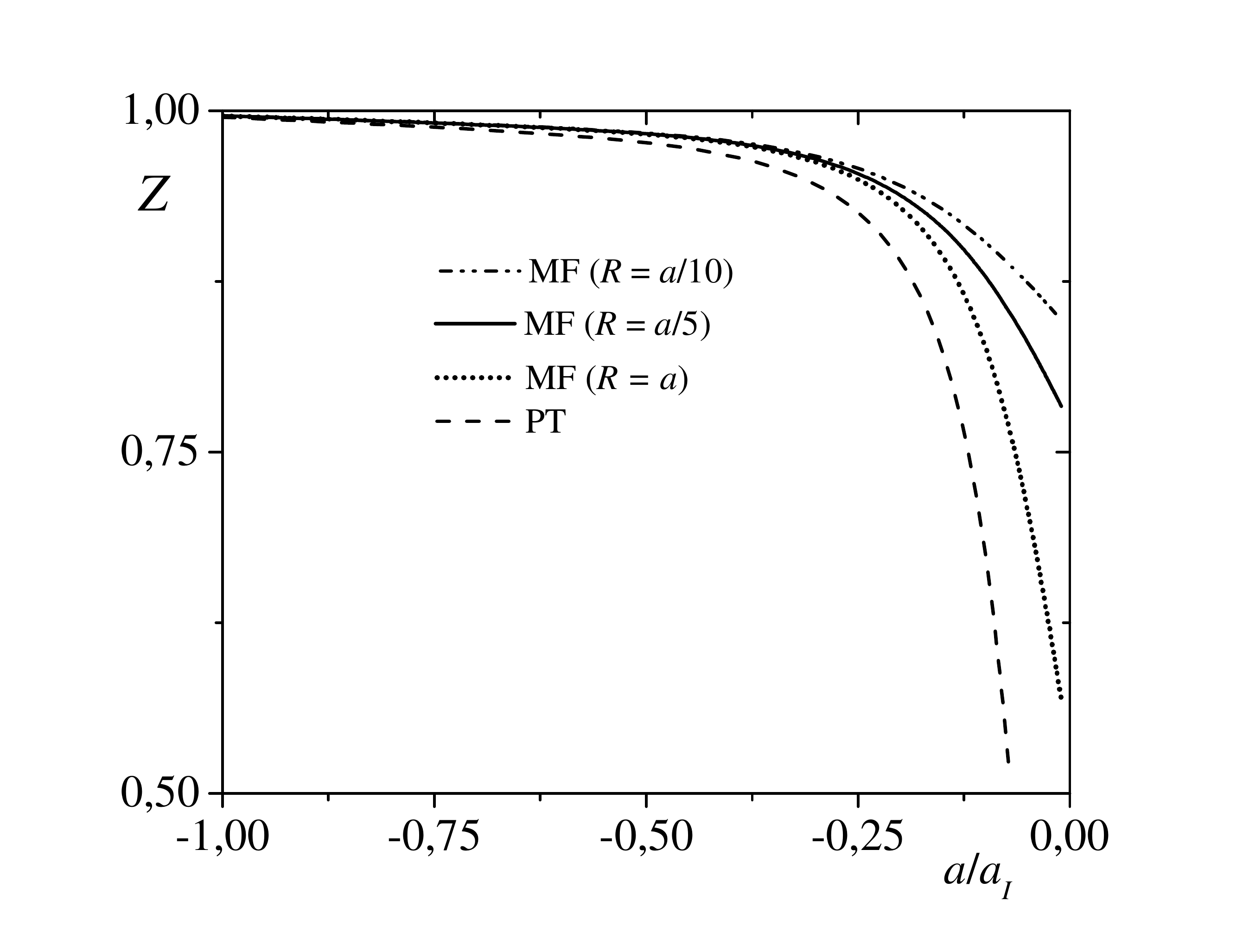}
	\caption{Effective mass and quasiparticle residue of the Bose polaron in dilute gas, $na^3=10^{-5}$. Solid, dotted and dash-dot-dotted lines display the present MF calculations at $R<a/5$, $R<a$ and $R<a/10$, respectively. The perturbative result up to the second order \cite{Grusdt_Demler_15,Christensen_15,Panochko_17} is shown by dashed line. Symbol is the quantum MC data from \cite{Pena_Ardila_15}.}
	\label{mass_MF}
\end{figure}
The results of numerical calculations for the MF binding energy, effective mass and quasiparticle residue of polaron immersed in a dilute Bose gas are presented in Figs.~\ref{energy_Mf}, \ref{mass_MF}. For comparison, we also plotted the results of the second-order perturbation theory and a quantum MC. Importantly that our MF calculations qualitatively agree with MC results in the whole range of the boson-impurity coupling including an unitary limit. It should be noted that MF does not reproduce the perturbation theory even at weak boson-impurity interactions, but provides the non-perturbative predictions in the strong-coupling limit. Actually, the standard perturbation theory results are clearly inapplicable  at unitary, because the binding energy and the effective mass diverge, while the quasiparticle residue falls down to negative values. Furthermore, the MF results (particularly the effective mass and the residue) are sensitive to the range of the boson-impurity potential in close vicinity of the unitary limit $a/a_I \to 0$. Another interesting consequence of this study is that, likewise the MC results, both the binding energy and the effective mass are finite at unitary limit signaling absence of the impurity self-localization phenomenon.

\subsection{Impurity in unitary Bose gas}
At unitary, the two-boson scattering amplitude diverges in the infrared region, and as a consequence, the many-particle system demonstrates universal behavior. Particularly, at absolute zero the energy density is given by $\mathcal{E}[n]=\xi\frac{3}{5}\frac{\hbar^2}{m}n^{5/3}$ with $\xi$ (in case of fermions it is often called the Bertsch parameter) being the true number. A cursory glance on the literature concerning the determination of parameter $\xi$ for bosons shows that there is no consensus at this point. Various theoretical approaches give quite different values, for instance, $22.22$ \cite{Cowell}, $5.0$ \cite{Lee} and $4.0$ \cite{Diederix}. For the numerical calculations, however, we used experimental result $3.40$ \cite{Navon} and the one obtained in the quantum MC simulations $5.32$ \cite{Rossi}.

\begin{figure}[h!]
	\includegraphics[width=0.45\textwidth,clip,angle=-0]{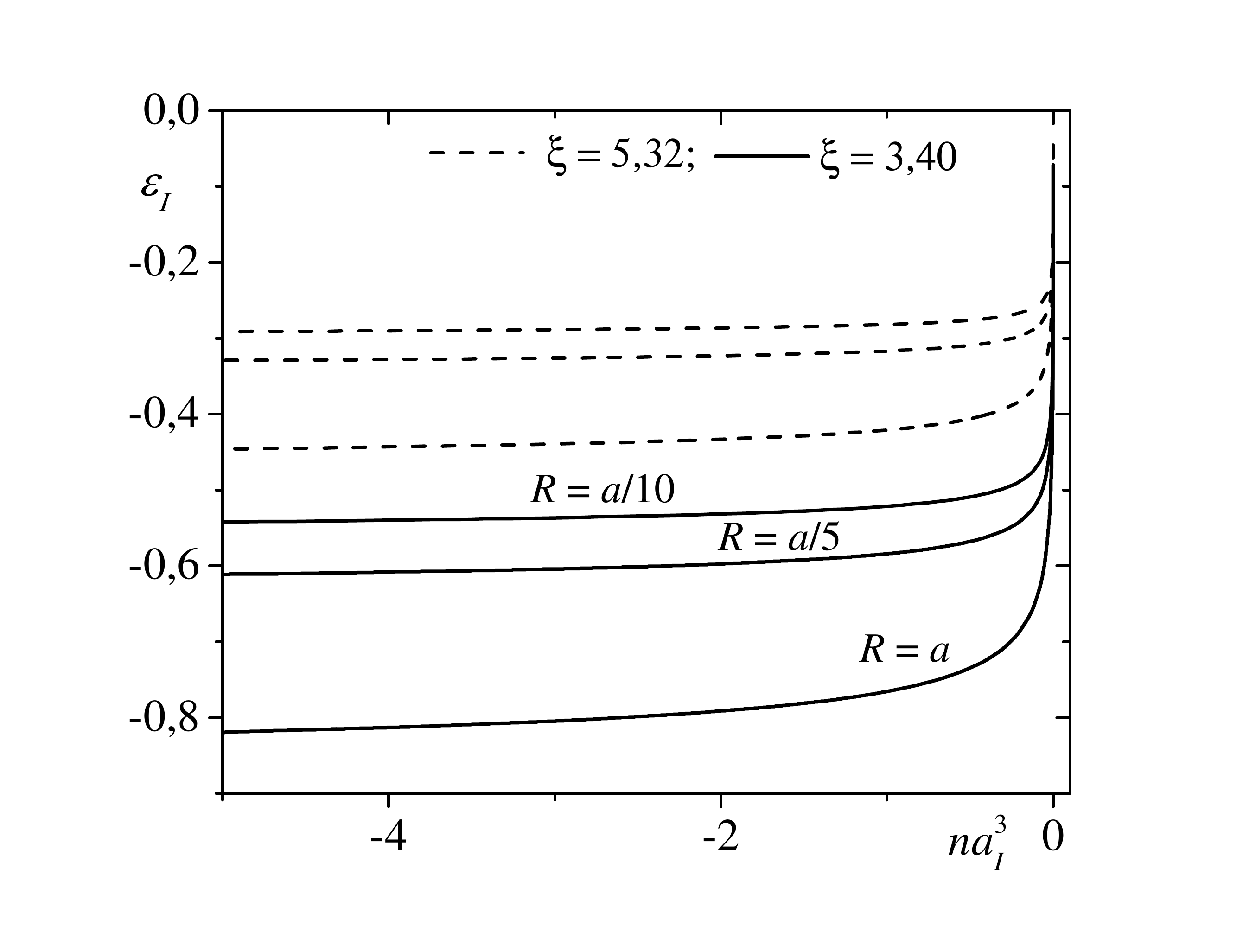}
	\includegraphics[width=0.45\textwidth,clip,angle=-0]{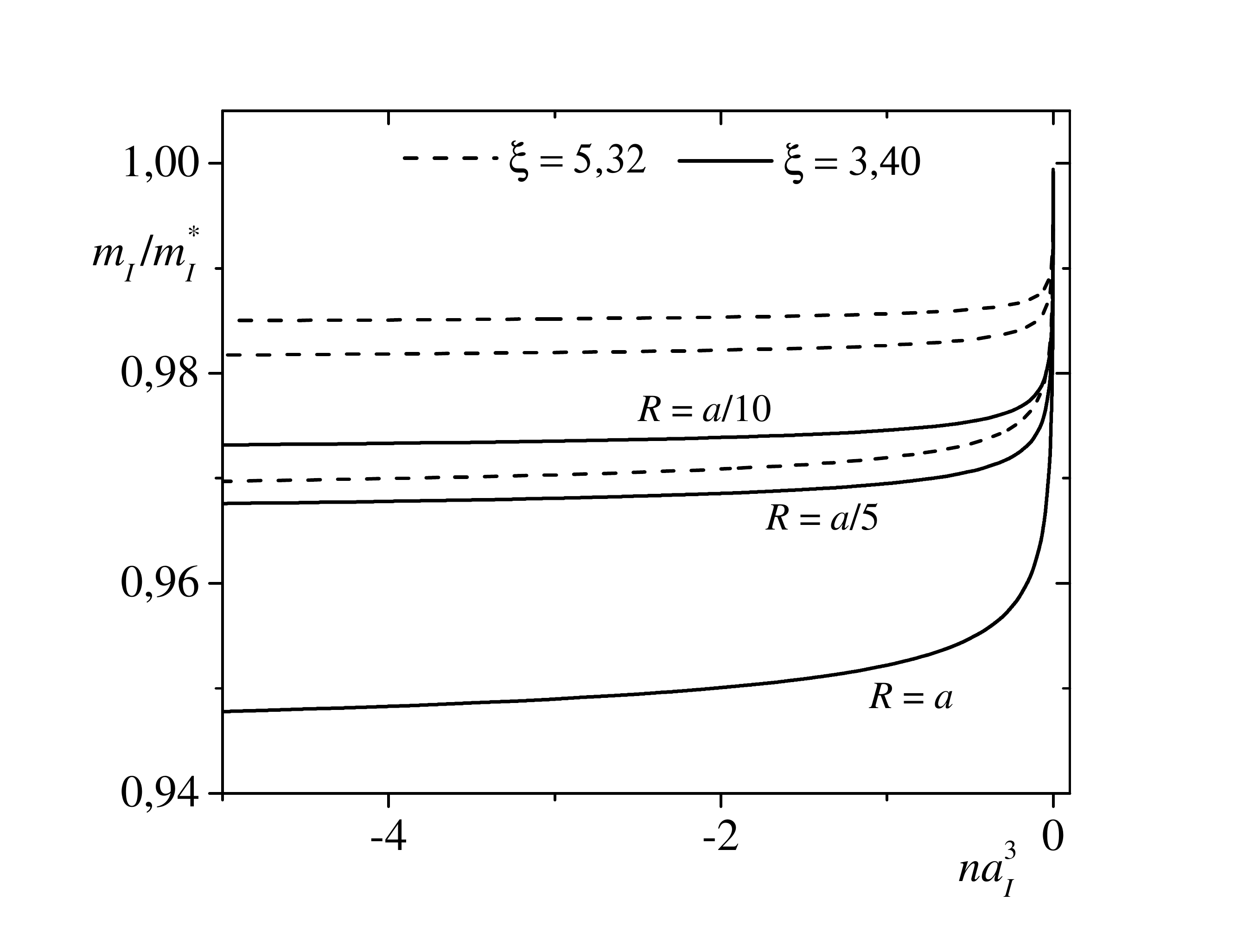}
	\includegraphics[width=0.45\textwidth,clip,angle=-0]{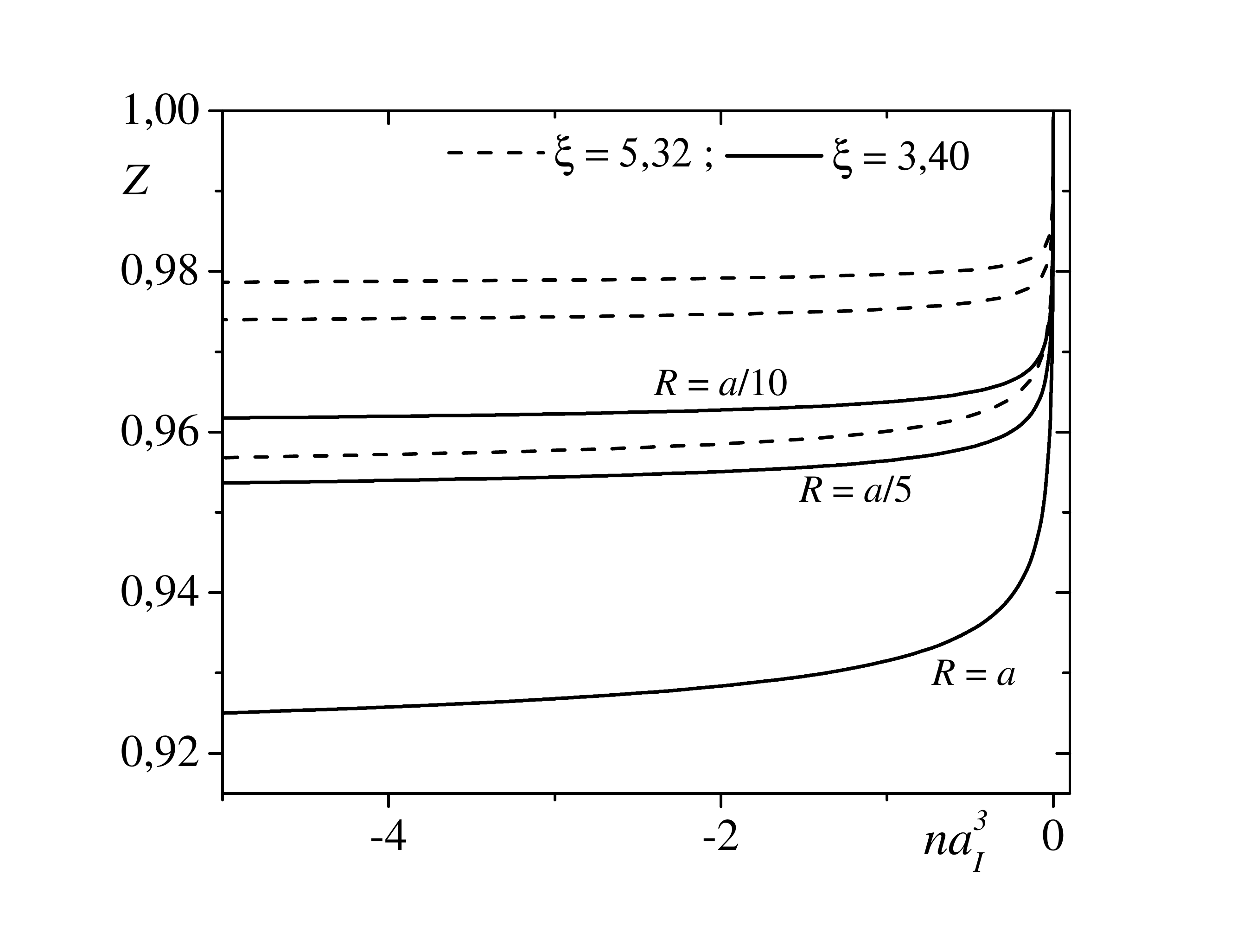}
	\caption{The binding energy (in units of $\xi\frac{\hbar^2}{m}n^{2/3}$), the inverse effective mass and quasiparticle residue of the attractive Bose polaron immersed in unitary gas. Solid and dashed lines represent results for different Bertsch parameters and ranges of potential (curves from top to bottom) $R=a/10$, $R=a/5$ and $R=a$.}
	\label{energy_LDA_minus}
\end{figure}

\begin{figure}[h!]
	\includegraphics[width=0.45\textwidth,clip,angle=-0]{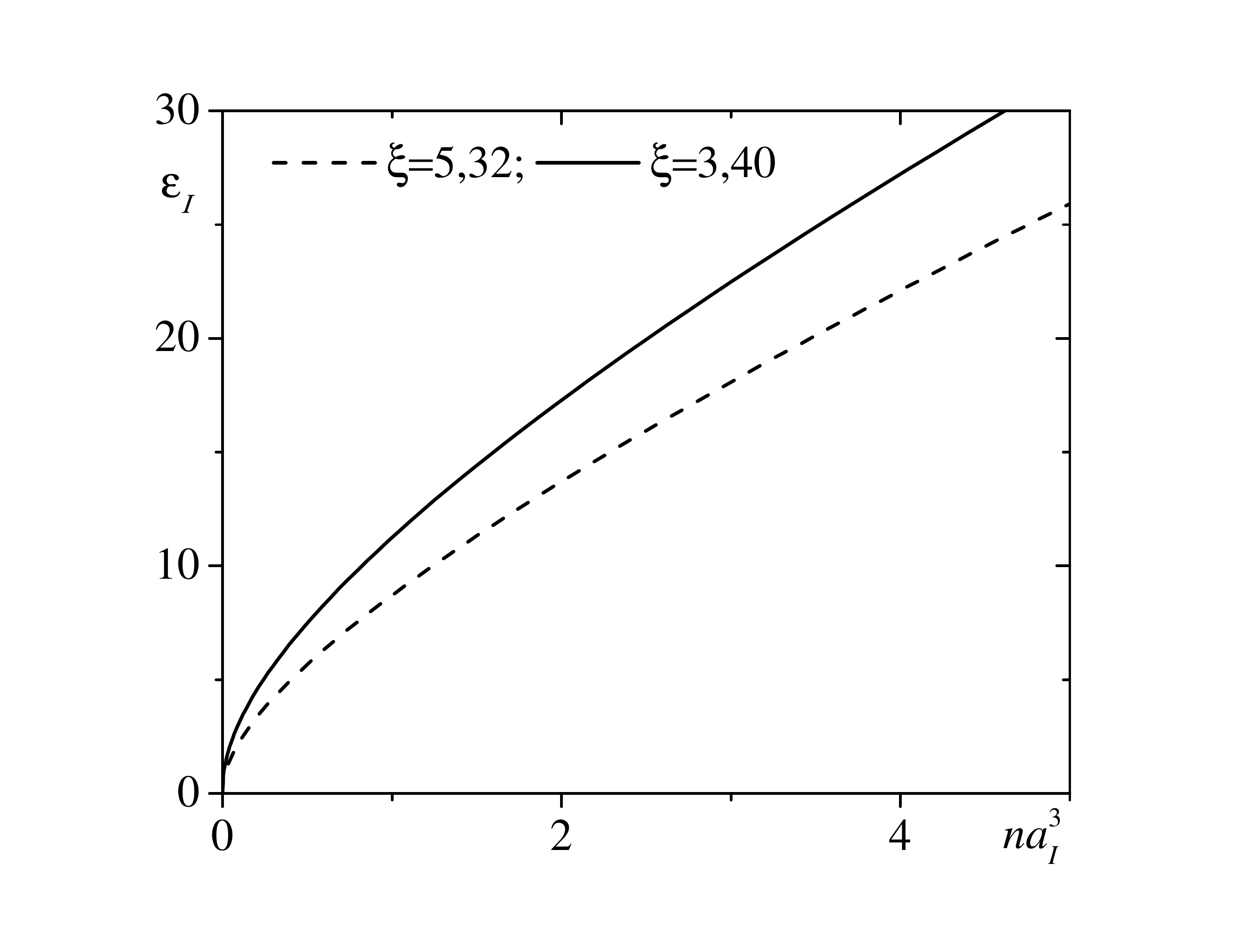}
	\includegraphics[width=0.45\textwidth,clip,angle=-0]{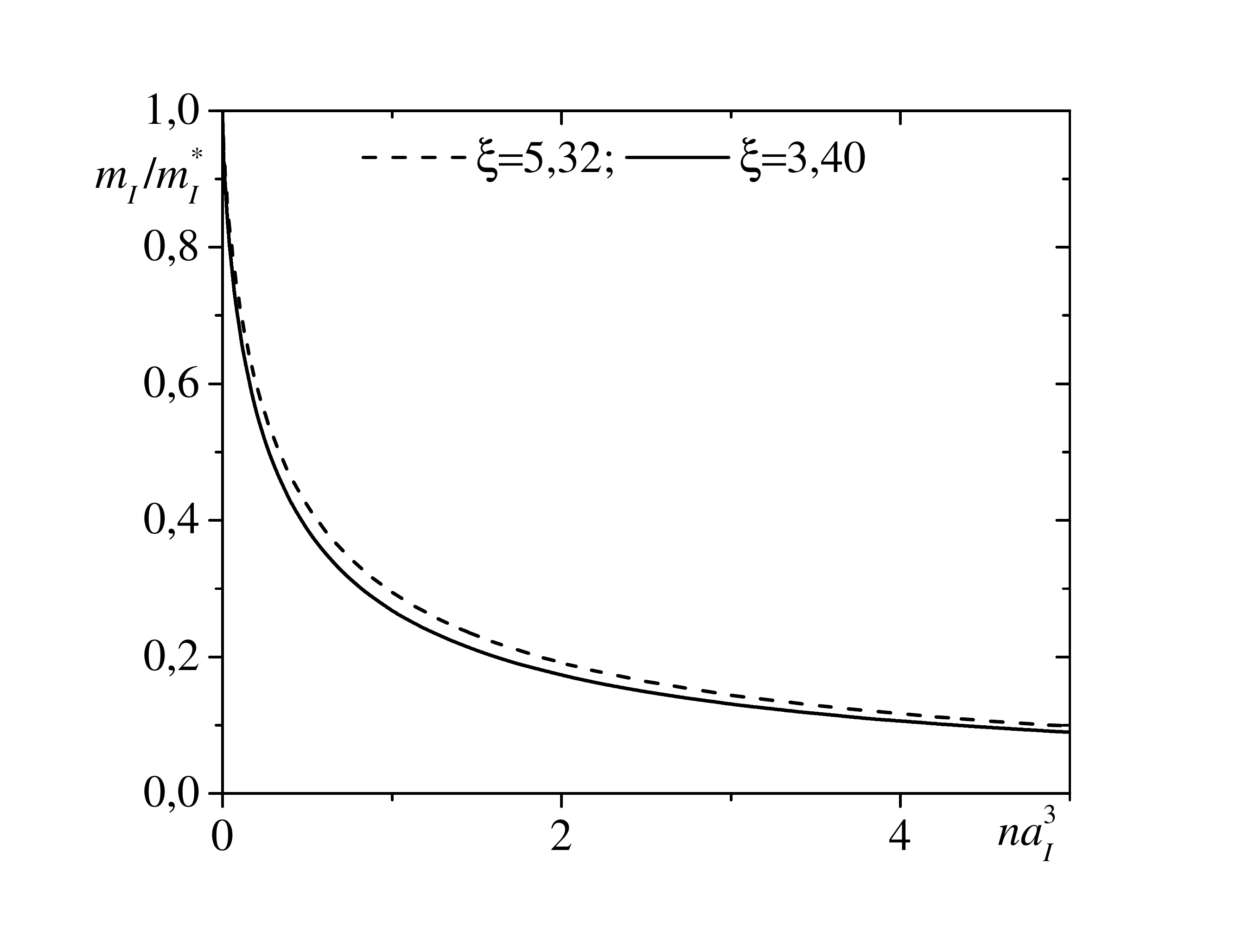}
	\includegraphics[width=0.45\textwidth,clip,angle=-0]{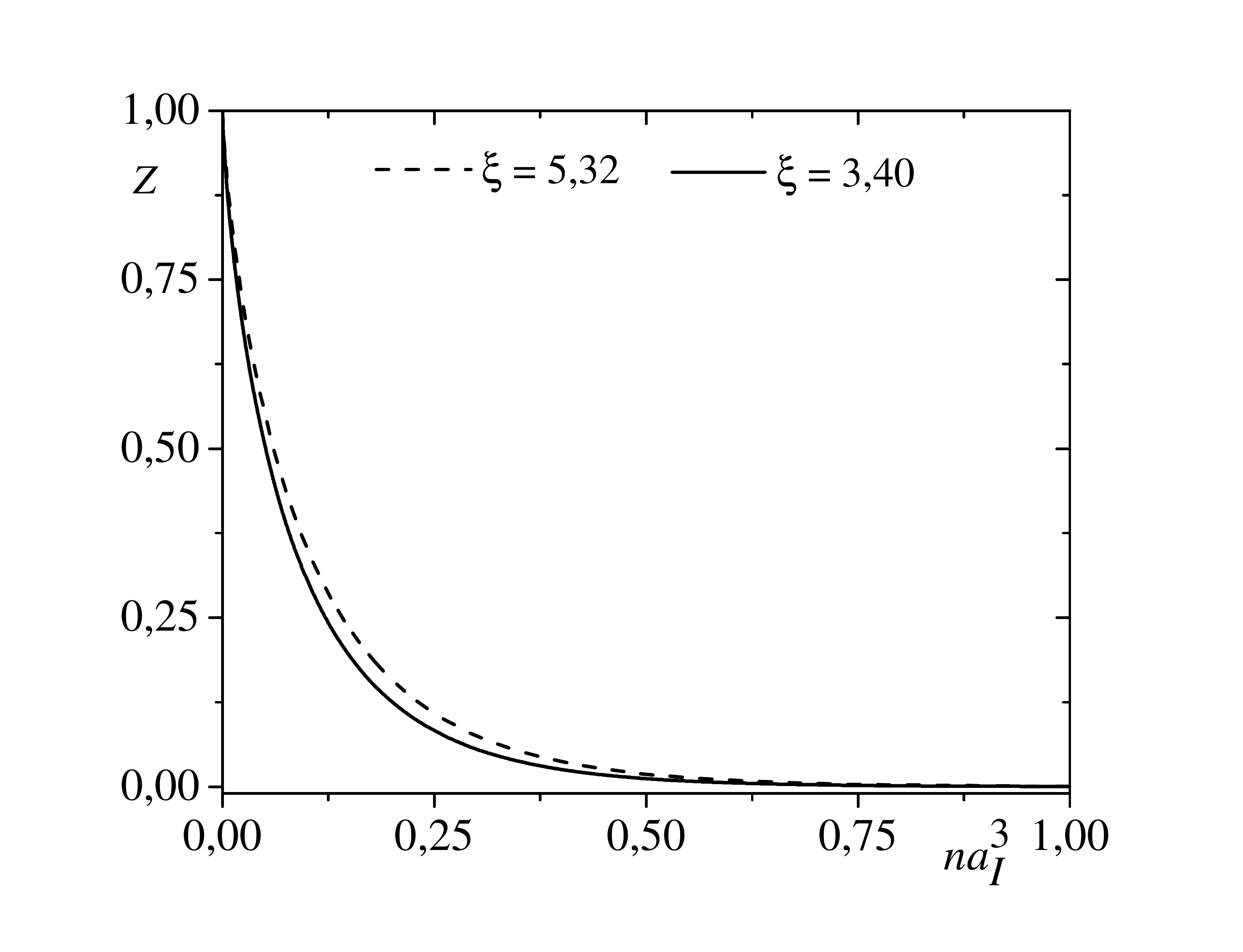}
	\caption{The binding energy (in units of $\xi\frac{\hbar^2}{m}n^{2/3}$), inverse effective mass and quasiparticle residue of the repulsive polaron immersed in unitary Bose gas. Solid and dashed lines represent results for different Bertsch parameters.}
	\label{energy_LDA}
\end{figure}

To study the attractive branch of polaron in unitary Bose gas we applied the numerical procedure to Eq.~(\ref{phi}) with $\mathcal{E}'[n]=\xi\frac{\hbar^2}{m}n^{2/3}$ and a two-body interaction (\ref{Phi}). We were interested only in real spherically-symmetric solutions for $\phi({\bf r})$ that satisfy two boundary conditions, $\phi(0)=0$ and $\phi(\infty)=\phi_{\infty}$, where the latter constant is related to the average density of bosons, an intensity of the boson-impurity interaction, a mass ratio and a value of the Bertsch parameter $\xi$. As previously, we took the same values of ranges and depths of the square-well potential and set $m_I=m$. Results for attractive polaronic branch are presented in Fig.~\ref{energy_LDA_minus}
For the repulsive Bose polaron we used the hard-sphere potential instead. The reason for this is that we could not find a unique stable solution of the non-linear differential equation (\ref{phi}) with the spherical-well boson-impurity interaction for positive scattering lengths. For the hard-sphere potential the $s$-wave scattering length $a_I$ coincides with the radius of hard core and mathematically the role of potential reduces to the change of boundary condition at the origin by the following one $\phi(a_I)=0$. The results for the binding energy, effective mass and quasiparticle residue as functions of dimensionless scattering length $a^{3}_In$ for the impurity in unitary Bose gas are presented in Fig.~\ref{energy_LDA}.
We see again that the self-localization does not occur for an impurity immersed in the strongly-interacting Bose gas, particularly, parameters of the low-momentum Bose-polaron spectrum remain finite even in the limit $a_I\to -\infty$. It worth recalling that the hard-sphere potential inadequately describes properties of the model at large positive $a_I$s, therefore, results for the repulsive branch should be applicable only in the weakly-interacting $na^3_I\ll 1$ regime. A very unexpected conclusion of these calculations is that, in general, a behavior of impurity (see Fig.~\ref{energy_LDA_minus}) is almost unaffected by the interaction with bosons, especially the effective mass $m^*_I$ and the residue $Z$, which at unitary are less than $10\%$ differ from their bare values. The MF approximation, however, provides only the lower bound for low-energy parameters of the impurity spectrum and results obtained here awaiting for confirmation by more sophisticated approaches.

\section{Summary}
In conclusion, we have explored, for the first time, properties of a single impurity atom immersed in unitary Bose gas. Particularly, by utilizing an original MF-like approach, which treats Bose system semi-phenomenologically and assumes that all bosons occupy the same one-particle state, and was shown to describe a behavior of Bose polaron in the dilute Bose gas on the qualitatively correct level, we have calculated the low-momentum parameters of the impurity spectrum, i.e., the binding energy, the effective mass  and the quasiparticle residue in a wide region of the boson-impurity couplings. Although, the experimental realization of such systems is complicated by the meta-stable nature of a unitary Bose gas prepared in the so-called `upper branch' \cite{Li_Ho}, it will be interesting to test our predictions in MC simulations. Another possible application of the presented approach is the problem of impurity in fermionic superfluids at unitary. The only modifications of the presented scheme required for such a calculations are the use of fermionic Bertsch parameter and $m$ should be identified with the doubled mass of Fermi particles.

\begin{center}
	{\bf Acknowledgements}
\end{center}
We are grateful to Dr.~L.~A.~Pe\~na Ardila for sharing with us results of Monte Carlo simulations. Work of O.~H. was partly supported by Project FF-83F (No. 0119U002203) from the Ministry of Education and Science of Ukraine.

\end{document}